# Hybrid Interface States and Spin Polarization at Ferromagnetic Metal-Organic Heterojunctions: Interface Engineering for Efficient Spin Injection in Organic Spintronics

Shengwei Shi[1,*], Zhengyi Sun[1], Amilcar Bedoya-Pinto[2], Patrizio Graziosi[3], Xin Li[4], Xianjie Liu[1], Luis Hueso[2], Valentin A. Dediu[3], Yi Luo[4], and Mats Fahlman[1,*]

[1]Department of Physics, Chemistry and Biology, Linkoping University, SE-58183 Linkoping, Sweden
E-mail: shesh@ifm.liu.se, mats.fahlman@liu.se
[2]CiC nanoGUNE Consolider, Tolosa Hiribidea, 76, E-20018 Donostia-San Sebastian, Spain
[3]ISMN-CNR, Via Gobetti 101, 40129 Bologna, Italy
[4]Division of Theoretical Chemistry and Biology, Royal Institute of Technology, SE-10691 Stockholm, Sweden

**Keywords**: organic spintronics, hybrid interface state, spin polarization, interface engineering

**Abstract**: Ferromagnetic metal-organic semiconductor (FM-OSC) hybrid interfaces have shown to play an important role for spin injection in organic spintronics. Here, 11,11,12,12-tetracyanonaptho-2,6-quinodimethane (TNAP) is introduced as an interfacial layer in Co-OSCs heterojunction with an aim to tune the spin injection. The Co/TNAP interface is investigated by use of X-ray and ultraviolet photoelectron spectroscopy (XPS/UPS), near edge X-ray absorption fine structure (NEXAFS) and X-ray magnetic circular dichroism (XMCD). Hybrid interface states (HIS) are observed at Co/TNAP interface resulting from chemical interaction between Co and TNAP. The energy level alignment at Co/TNAP/OSCs interface is also obtained, and a reduction of the hole injection barrier is demonstrated. XMCD results confirm sizeable spin polarization at the Co/TNAP hybrid interface.

## 1. Introduction

Organic spintronics, a fusion of organic electronics and spintronics, represents a new research field,[1] where organic semiconductors (OSCs) are used to mediate or control the spin polarized signal as they consist mainly of atoms with low atomic number Z, leading to a low spin-orbital coupling and thus to extremely long spin relaxation times.[2, 3] In addition, the wide range of available chemical functionality of OSCs and the ability to precisely tailor their electronic (and optical) properties hold the potential to realize the efficient manipulation of electron spin. Though advances in device performances have been made in the past few years,[4, 5, 6, 7] fundamental knowledge of spin injection at the ferromagnetic metal (FM)/OSC heterojunction is still incomplete.[2] Nevertheless, FM/OSC hybrid interfaces represent a promising and intriguing material system in organic spintronics.[8]

To realize spin polarized injection, there are three important factors to be considered during the design of FM/OSC hybrid interfaces. Firstly, the carrier injection barrier, typically defined as the energy difference between the highest occupied molecular orbital (HOMO) or lowest unoccupied molecular orbital (LUMO) and the Fermi level ($E_F$), must be minimized by tuning the energy level alignment at the electrode interfaces, which have already been successfully realized in traditional organic electronics. For example, it has been demonstrated that the carrier injection barrier at the metal-organic interface can be tuned by use of a self-assembled monolayer,[9] or electron acceptor molecules as a buffer layer.[10] Although similar work has been done in organic spintronics to improve electron injection,[11] there are additional effects such as influence of the interfacial layer on magnetic moments of the FM electrodes and possible spin polarization of the OSC orbitals. The role of interface engineered vacuum level shifts on spin injection and extraction is thus a topic of high current interest.[6, 12, 13] Secondly, the conductance should match with each other for FMs and OSCs at the interface. Although the OSCs with high mobility are selected to use as the space layer,[14] the so-called conductance mismatch is a general probelm in organic spintronics because of huge difference in the conductance of FMs and OSCs.[15] In this case, the spin properties of the interface are dominated by the high resistance of OSCs, which is a non-magnetic molecule without spin discrimination. One successful method in organic spintronics is to use a tunneling barrier as spin filter, in which the mechanism of conductance mismatch does not apply because the transport is due to tunneling mechanism and not diffusion,[6] and also the OSC films can be thin enough so that the transport across the film is through multi-step tunneling.[16] Organic spin valves also have been successfully reported featuring a thick (>10 nm) OSC layer without a tunneling barrier and operated at a few millivolts bias voltage,[4] somewhat surprising given the conductivity mismatch likely present. For example, $Alq_3$ is the most popular OSC for organic spin valves, and the spin diffusion length was determined to be 45 nm at 11 K,[4] but it has been debated for the reproducibility of the published experimental results,[17] and also there are debates on which carrier transport in $Alq_3$ layer and the polarity of magnetoresistance (MR).[12, 18] Early reports suggest spin transport can only be seen through pinholes in the $Alq_3$ layer,[19] while more recent work has shown that $Alq_3$ can be used for effective spin transport by using a low work function metal.[20] Although there are some controversies on spin transport in thick layer of $Alq_3$, it is worth stressing that MR in manganite (LSMO)/$Alq_3$/Co injection devices is an established result that is corroborated by several experimental reports.[2]

Besides the above two points concerning about the "carrier injection" at the interface, the third factor to be considered is "spin polarization" or spin filter effects. A pristine nonmagnetic OSC is spin undiscriminate because it features the same spin-up/down resistances. When an OSC is put in contact with an atomically clean FM surface, hybridization between 3d electrons and molecular orbitals may take place at the interface from the chemisorption of the OSC on the FM surface, which results in the formation of so-called hybrid interface states (HIS).[21, 22] The electronic structure in this case might be modified to show new density of state (DOS) near the $E_F$, arising from the new HIS. This new DOS can determine the spin-polarization of the injected current, which then can be dramatically different, and even reversed, compared with the polarization of the electrodes.[21, 23] As HIS are generated just at the interface of the first monolayer of OSCs and the top layer of FM,[22] it acts as spin filters at the hybrid interface,[8, 21, 22, 24] or one can say it works as an interfacial layer between FM and OSC layer, and the selection of appropriate molecule as subsequent layer to couple with this HIS can ensure spin polarization in a spin conserving way. Another idea is to use a specific OSC molecule to induce a desired HIS at the FM interface, followed by deposition of a (different) OSC film with desired film forming and transporting properties. Such an approach provides a larger flexibility in tuning the energy level alignment and spin polarization/filtering at the interface while obtaining sufficient quality in film uniformity and charge/spin transport.

The modification of the binding nature at FM/OSC hybrid interfaces represents the main means for the tailoring of the spin transfer behavior in organic spintronics.[12, 23] One of the simplest approaches is to use nonmagnetic dipole molecules as an interfacial layer to modify the binding and induce a dipole energy shift between FM and OSCs.[12] By inserting a dipole molecule layer at FM/OSC hybrid interface, the spin transfer efficiency may be modified by leading to resonances between LUMO&HOMO and spin polarized metal density of state. This provides an appealing way of engineering spin-selective injection channels at FM/OSC interface by proper selection of dipole molecules.[25]

11,11,12,12-tetracyanonaptho-2,6-quinodimethane (TNAP) is an electron acceptor molecule,[26] which has shown strong interaction with clean metal surfaces forming hybrid interfaces.[27] Moreover, tetracyano-based molecules are promising materials to form room temperature metal-organic magnets, resulting from the interplay between the metals (such as V and Ni etc) and the tetracyano-based ligands,[28] hence it is foreseeable to realize the spin injection and transport at organic-inorganic interface.[29] Thus, TNAP shows potential as an interfacial layer in organic spintronics. In this work, we investigate the interface of TNAP on Co by use of X-ray photoelectron spectroscopy (XPS), ultraviolet photoelectron spectroscopy (UPS), near edge X-ray absorption fine structure (NEXAFS) and X-ray magnetic circular dichroism (XMCD). The energy level alignment at Co/TNAP/OSC interfaces is obtained and the magnetic properties of the hybrid FM-organic interface are determined, providing information on the key properties controlling spin-polarized injection.

**2. Results and Discussions**
**2.1. Interface Energetics**

In this section, we introduce experimental results on the energy level alignments related to different heterjunctions between Co and organic molecules. In the first step, we investigate the Co/TNAP interface to understand the modification of TNAP on the electronic structure of Co and the possible interaction between Co and TNAP. As a second step, we investigate the function of TNAP as an interfacial layer in real device structures, by obtaining the energy level alignment over FM/TNAP/OSCs trilayer stacks, where tris(8-hydroxyquinolinato) aluminium (Alq$_3$) and C$_{60}$ are chosen as the OSCs given their frequent use in organic spintronic devices.[4, 6, 7]

*2.1.1. Co/TNAP*

UPS spectra of TNAP deposited on Co substrate are shown in **Figure 1**, where the TNAP thickness range from roughly a monolayer (~ML) to multilayer (clean bare Co surface is also shown as a reference). In Figure 1(a), the secondary electron cut-offs (SEC) provide the evolution of the work function. By subtracting the binding energy of the SEC from the excitation energy (21.22 eV), the work function of the as-deposited Co film is found to be about 5.0 (±0.1) eV, in good agreement with previously published values.[30] The work function shows an increase from 5.0 eV for bare Co film to 5.7 eV when a ~ML on Co is deposited, whereupon it gradually decrease at increasing thickness until it saturates at 5.2 eV for the thick TNAP layer. This suggests that TNAP layers can reduce hole injection barrier and improve hole injection when it is used in real device and offers the possibility to tune the barrier height based on the TNAP interface layer thickness.[31] In Figure 1b, the valence band spectrum show eight main spectral features at the Co/TNAP interface: A (0.25 eV), B (1.1 eV), C (2.5 eV), D (3.3 eV), E (4.7 eV), F (6.1 eV), G (6.9 eV) and H (8.7 eV). From Figure 1(c) it is evident that with TNAP thickness increasing, the intensity for both A and B decreases and finally disappears for "bulk" TNAP. A and B hence are assigned as hybridized interface states (HIS) caused by the interaction between Co and TNAP, similar to the case of Alq$_3$ on Co.[18] The results also are similar to the reports on Cu/TNAP where Cu is considered to diffuse into TNAP at the interface and form charge transfer complex with TNAP.[27] Peak C corresponds to the HOMO of TNAP and it becomes more prominent as film thickness increases to bulk. The intensity of peak D (3.3 eV) increases with the thickness of TNAP, and it was not observed in Cu/TNAP but only for thick film above 0.38 nm in Ag/TNAP,[27] suggesting that unlike Cu, Co does not diffuse into the whole TNAP film and thus pristine TNAP layers are formed away from the interface. In Figure 1(c), the Fermi edge is still clear even when TNAP shows a nominally thick film, which means TNAP is not uniform and forms in the island mode on the Co surface. In this case, although the first layer of TNAP has a strong interaction with Co by charge transfer, it only happens on partial Co surface, and subsequently deposited molecules can fill into other empty Co surface to continue the chemisorption until the bare Co surface is fully covered by TNAP molecules. After that, the following deposited molecules stick on the chemisorbed TNAP layer (where TNAP is close to TNAP$^-$), and TNAP$^0$ overlayers are subsequently formed. Finally, the peaks for E, F, G and H are considered to be main features for neutral TNAP, which show similar results with the calculation by K. Tanai *et al*,[27] and they also show increasing intensity with the thickness of TNAP layer.

Based on the results and above discussion, the energy level alignment for the Co/TNAP hybrid system is given in **Figure 2**, and HIS is shown as red rectangle near $E_F$ as well (Figure S1).

*2.1.2. Co/TNAP/Alq$_3$*

In **Figure 3**, we present the UPS spectra for the multilayer stack of Co/TNAP/Alq$_3$ where the TNAP is roughly a ML thick. As is well known, Alq$_3$ is a very sensitive molecule to investigate the energy level alignment because of its strong intrinsic dipole moment,[32, 33] and a fully ordered Alq$_3$ layer could result in a shift of about 1 eV.[34] Because we mainly care about the interfacial effects, here we only discuss the effects of first few layers of Alq$_3$ on the Co/TNAP interface.

From Fig 3a, with ~ML TNAP adsorption on Co surface, the work function changes from 5.0 eV to 5.7 eV. Upon deposition of a very thin Alq$_3$ layer on ~ML TNAP, the work function is reduced by 0.5 eV. There are no new features appearing in the near $E_F$ region upon Alq$_3$ deposition (Fig 3b), only the two clear gap states (peak A and B) existing near $E_F$ and two spectral features C and D, all belonging to the Co/TNAP interfaces as we have discussed in the above section. The work function gradually decreases to 4.5 eV in the thickness range of our measurements (Fig 3a), with a 1.2 eV reduction comparing with that of Co/TNAP. The big interface dipole at the trilayer stack may originate from the charge transfer from Alq$_3$ to TNAP[32]. In addition, because the ~ML TNAP likely does not fully cover the Co surface (Figure 1c), the reduced work function may also result from Pauli repulsion which is expected to work at partial Co/Alq$_3$ interfaces.[35] The HOMO feature of Alq$_3$ (peak A') in Fig 3(b) appears and intensifies upon increased Alq$_3$ deposition while at the same time the spectral features (Peak A, B, C and D) existing at Co/TNAP interface are gradually suppressed and finally disappear (Fig 3b). The HOMO edge is situated at 1.1 eV, while it is at 2. 1 eV for Co/Alq$_3$ without the TNAP interfacial layer:[30] a 1.0 eV shift.

Based on the UPS measurements, the energy level alignment for Co/TNAP/Alq$_3$ is given in **Figure 4**(a). The hole injection barrier is determined by the gap of $E_F$ and HOMO. Comparing with the results of Co/Alq$_3$ interface before,[30] the hole injection barrier from the hybrid interface to a certain thick layer of Alq$_3$ is reduced by 1.0 eV when ~ML TNAP is inserted between Co and Alq$_3$, which means that TNAP as interfacial layer can improve the hole injection when hole is the dominant carrier in spintronics based on Alq$_3$ barrier.[12]

*2.1.3. Co/TNAP/C$_{60}$*

In Figure 3c and d, we show the UPS spectra for hybrid interface of Co/TNAP/C$_{60}$ where TNAP is a ~ML thick. From Fig 3a and c, with ~ML TNAP adsorption on Co, the work function is enhanced by 0.7 eV comparing with that of bare Co, and the TNAP (ML) yields reproducibly the work function of ~5.7eV. The subsequent deposition of a thin C$_{60}$ layer reduces the work function from 5.7 eV to 5.3 eV, which gradually satruates at 5.1 eV for thick C$_{60}$ layer with 0.6 eV shift comparing with that of Co/TNAP interface. The deposition of C$_{60}$ layer greatly changes the electronic structure in valence band, and the gap states (peak A and B) existing at the Co/TNAP interface are gradually suppressed and finally disappear for thick C$_{60}$ layer. The same situation happens for the spectral features C and D in ~ML TNAP (Fig 3d). The spectra features A' and B' for C$_{60}$ appear immediately even when a very thin layer is formed on TNAP, and because they are so strong, the spectra features for TNAP on Co show a very low intensity. In addition, the features for C$_{60}$ are very stable in binding energy with thickness increasing, but they become more and more narrow and show increasing intensity. The HOMO edge is situated at 1.2 eV, while it is at 1. 5 eV for Co/C$_{60}$ without the TNAP interfacial layer (see supplement): a 0.3 eV shift.

Based on the UPS measurements, the energy level alignment for Co/TNAP/$C_{60}$ and Co/$C_{60}$ is given in Figure 4b and Figure S2c. When comparing the two heterjunctions with and without ~ML TNAP, we find that the hole injection barrier decreases by 0.3 eV from the interface to $C_{60}$ when TNAP layer is inserted between Co and $C_{60}$, which means that a ~ML TNAP as interfacial layer can improve the injection when holes act as spin carriers in organic spintronic devices based on $C_{60}$ barrier.

**2.2. Synchrotron Techniques**

NEXAFS and XMCD are two powerful methods in synchrotron radiation techniques, which have shown important roles in nanomagnetism and spintronics.[36] With NEXAFS, it's possible to extract information on unoccupied molecular orbital states, and molecular orientation on substrates, and the most importantly, it is element-specific and polarization dependence. XMCD is based on the difference between the NEXAFS spectra taken by left and right circular polarized incident light, by which the spin and orbital momentum in FM atoms can be determined. In this section, firstly, we introduce the NEXAFS measurements for Co/TNAP hybrid interface and analyze the difference between ~ML and multilayer TNAP on Co surface combining with theoretically calculations, in the next step, we dissucss the orientation of ~ML TNAP on Co surface by angle-dependence NEXAFS. Finally, we introduce the XMCD results on N K-edge and Co L-edge, and we investigate the effect of ~ML TNAP adsorption on the magnetic properties of Co by use of XMCD sum rules[37].

*2.2.1. NEXAFS*

**Figure 5** shows NEXAFS spectra of N and C K-edges in both multilayer and ~ML TNAP on Co, recorded with a photon line width of about 100 meV, and the calculated results on gas-state TNAP are also given as the reference. The optimized structure of TNAP from the calculation is given in Figure 5a to show the elements in different chemical environment. We can see that both calculated C and N K-edges have good agreement in spectral profiles comparing with experimental results. However, the peak positions and the peak separations are different. This is probably because the calculated spectra are based on a single TNAP molecule, while the samples measured in experiment are films, thus inducing intermolecular screening effects.

There are clear features in the calculated N K-edge spectrum (Figure 5b). The first peak is situated at 396.7 eV corresponding to the excitation into the LUMO, with the next two peaks at 398.5 and 399.7 eV corresponding to excitations into LUMO+1 and LUMO+2. Since the four N atoms in TNAP molecule have a near identical configuration, the resulting spectra are quite similar and have almost the same weight of contribution to the final NEXAFS spectrum. Final state molecular orbital (MO) calculation from equivalent core hole approximation (ECH) shows that all of the first three peaks are 1s to $\pi^*$ features.[38] The LUMO+1 MO (second peak) is localized around the excited N atom with small delocalized feature, while the LUMO and LUMO+2 (first and third peaks) MOs are more delocalized than LUMO+1. In the experiment results, LUMO feature around 396.7 eV shows a weak but wide peak for ~ML TNAP compared with the multilayer film, likely a result from the charge transfer from the Co into the TNAP LUMO at the interface that is evident from the UPS measurements. There are nearly no energy shift in LUMO+1 for ~ML and multilayer, but it shows weak peak-broadening in ~ML TNAP. Combining with XPS results of N 1s (Fig. S3b), it seems that the chemical interaction between Co and TNAP happens through the bonding of Co to the N-atoms (see the supplement). It also should be pointed out that there is a shoulder peak at 395.9 eV in ~ML TNAP on Co due to Co second-order L-edge (inserted in Fig 5b), which is suppressed in multilayer TNAP.

The calculated C K-edge is more complex than the N K-edge. There are many peak features (Figure 5c), but the peak at 285.4 eV has a relative high intensity than the others. Calculated spectra for each individual C atom show that this peak is almost exclusively derived from the C atoms that bond with N, and the same for the peak around 286.7 eV, *i.e.*, these peaks are related to excitations involving such sites. The onset of absorption occurs by the creation of a core hole at the C5 atom and excitation into the LUMO. Final state MO calculation for C atom shows that, similar to the N atom, the resulting peaks at 285.4 and 286.7 eV are also 1s to π* resonances, corresponding to final LUMO+1 and LUMO+2 states, while LUMO+1 is more localized than LUMO+2.

Now we discuss the experimental results of the C K-edge (Figure 5c), which as noted is much more complicated than N K-edge because of the eight nonequivalent C atoms in the TNAP molecule. There are small but clear features in the frontier orbitals for multilayer TNAP in good agreement with the calculations, as this film is more molecule-like. There is a clear difference on the frontier orbitals in NEXAFS spectra between multilayer and ~ML TNAP, however. Comparing with the multilayer, several small features in low photon energy region, which are denoted by the green rectangle in Figure 5c, disappear in the ~ML film. There are nearly no shift in the photon energy for the main features (LUMO, LUMO+1 and LUMO+2), which also suggest that the interaction between the TNAP and Co is facilitated through the nitrogen boding to the Co-surface. Overall, the modification of the C and N K-edges spectra from ~ML to multilayer suggests mainly charge transfer and comparatively weak interaction with the carbon atoms, as compared with other molecules on Co.[39] The present NEXAFS results confirm the conclusion in UPS measurements that strong interaction exists at the TNAP/Co interface through bonding with the nitrogen atoms, charge transfer occurs from Co to TNAP and the frontier unoccupied states are (weakly) hybridized.

It is reasonable that the hybridization is correlated to the orientation of the chemisorbed molecules on the FM surface.[40] **Figure 5d** shows the angle-dependence NEXAFS for C K-edge in ~ML TNAP on Co, in which $\alpha$ means the angle between the incident X-ray and sample surface. As the angle $\alpha$ increases from 10 to 90 degree, the intensity in the 280-290 eV region (corresponding to C 1s→$\pi^*$) declines relatively compared with that in the 290-300 eV region (corresponding to C 1s →$\sigma^*$). Based on the experiments, the ~ML TNAP molecule prefers to lie (nearly) flat on the Co surface.

*2.2.2. XMCD*

UPS and NEXAFS results demonstrate the existence of HIS and strong interaction at Co/TNAP interface. We now examine the potential spin polarization in HIS induced by the FM substrate, which have been shown to change the spin-selective injection channels at $E_F$ of the hybrid interface and engineer spin injection in organic spintronics.[22] Recent reports have already shown that the hybridization between molecular orbitals and FM 3d valence band states can lead to induced magnetization and sizeable interfacial magnetic moment in OSCs.[18, 39, 41] It is another interesting question whether such strong hybridization significantly affects the spin-dependent electronic structure of the FM surface atoms. If so, one would expect to observe a change in the orbital and/or spin moments from Co L-edge XMCD spectra upon adsorption of TNAP onto Co.

As is well known, Co is a good spin injector because of its high spin polarization at the Fermi level, and it has important applications in organic spintronics.[4, 5, 6, 7] Unfortunately, it shows a strong signal due to second-order L-edge at the photon energy of 388.6 eV ($L_3$) and 395.9 eV ($L_2$) in NEXAFS/XMCD measurements,[42] which makes it difficult to confirm (possible) XMCD features from the N K-edge.[43] **Figure 6** shows XMCD spectra taken in TEY mode for the FM/TNAP interface (~ML), in which the XMCD curves are obtained by subtracting left-hand polarized NEXAFS spectra with spectra taken with right-hand polarized light. It's very clear that there are two peaks in the XMCD results denoted by A and B (Figure 6a). Peak A is broad, likely resulting from the overlap of the signals from Co second-order $L_2$-edge and N K-edge. However, peak A shows a small component at around 396.7 eV, besides the main one at 395.9 eV attributed to the Co L-edge second order, which should come from N K-edge in TNAP. As we have discussed in Figure 5b, the main feature for TNAP sits at 399.0 eV, and there ia a clear and sizeable XMCD feature at this photon energy (peak B), which can be the direct evidence of induced magnetization (spin polarization) in the TNAP molecule at the Co interface. Co L-edge XMCD signals are given to show the difference for the same sample before and after the adsorption of TNAP overlayers (Figure 6b). The spectra are normalized to the $L_3$-peak height of the NEXAFS sum spectra for parallel and antiparallel alignment between the magnetization and photon helicity. After the deposition of TNAP, the Co $L_3$ edge XMCD signal is slightly reduced while the $L_2$ edge remains unaffected, indicating a reduction in the magnetic moment on the Co atoms due to the interaction between Co and the TNAP molecules.

The so-called XMCD sum rules can be used to calculate the spin and orbital magnetic moments in FM metals,[37, 44] which makes it possible to obtain the quantitative influence of adsorbed TNAP on magnetic property of Co. According to the sum rules, we can obtain $\mu_S$ and $\mu_L$ from the integrals of the NEXAFS and XMCD spectra as:

$$\mu_S = -(6p - 4q)(10 - n_{3d})/r \qquad \mu_L = -4q(10 - n_{3d})/3r \qquad (1)$$

Where $n_{3d}$ is 7.51 corresponding to the number of Co 3d electrons calculated theoretically;[37] $p$, $q$, and $r$ are the integrated areas of the XMCD and NEXAFS sum spectra as defined in **Figure 7**.

The XMCD spectra are corrected by taking into account the incident angle (10° with respect to the sample surface) and the degree of circular polarization (85%), by multiplying the measured spectra by [1/cos(10°)]/0.85, while keeping the sum spectra the same. All the calculation results are listed in **Table 1**.

The magnetic moment obtained for the pristine Co sample (Figure 7a and 7b) are $\mu_S$ = 0.788 $\mu_B$ and $\mu_L$ = 0.101 $\mu_B$. These values are considerably smaller than the saturation values for Co as we expect. We think there are two possible reasons. First, the Co film is grown on amorphous substrate (Au-coated Si), and it should be in polycrystalline structure. In addition, the magnetic field (300 Oe) available at the beamline might not be enough to saturate the magnetization of Co film, and then the remnant magnetization is likely smaller than the saturation magnetization of Co. The orbital/spin ratio ($\mu_L/\mu_S$) for Co film shows a much higher value of 0.128 than the 0.095 in the bulk,[37] which is reasonable. For ultrathin 3d transition metal films, this ratio is typically enhanced, due to film-substrate d-orbital interaction and lifting of the orbital degeneracy by symmetry reduction at the surface.[37, 45]

The NEXAFS and XMCD spectra recorded at the Co $L_{2,3}$ edge in Co/TNAP are shown in Figure 7c and 7d. Using eq 1, we obtain $\mu_S$ = 0.782 $\mu_B$ and $\mu_L$ = 0.058 $\mu_B$, leading to a ratio $\mu_L/\mu_S$ = 0.074. Compared to the results obtained for the pristine Co substrate, $\mu_S$ is reduced by less than 1% (0.8%), indicating that the TNAP adsorption has nearly no harmful effect on the magnetic properties of Co, which should be advatageous for device application. The orbital magnetic moment is reduced by 42% because $\mu_L$ is more sensitive to hybridization effects.[39] However, $\mu_L$ is much smaller than $\mu_s$, thus it does not have a strong influence to the total magnetic moment. The results here provide a quantitative description of the effects of TNAP adsorption on magnetic moments of Co element in the surface, which is also in qualitative agreement with our observations of HIS and induced magnetization in N K-edge at Co/TNAP interface.

**3. Conclusions**

In summary, we concentrate on the function of TNAP as an interfacial layer in organic spintronics. We have examined the hybrid interface of Co/TNAP by use of photoelectron spectroscopy (XPS, UPS) and synchrotron techniques (NEXAFS, XMCD), and the trilayer stacks of Co/TNAP/OSCs are also investigated to obtain the information of energy level alignment in real device structure. The adsorption of TNAP on Co increases the work function of the interface, creating a better hole-injector from an energetics stand-point. As an interfacial layer between Co and OSCs, TNAP thus can reduce hole injection barriers and improve hole injection, which is promising for spintronic devices featuring holes as the (majority) spin-carriers. In addition, TNAP chemisorbs on Co inducing HIS near $E_F$, which is expected to result in the broadening of density of state and spin splitting to creat spin polarization at the hybrid interface. Sizebale spin polarization in N K-edge in TNAP is confirmed by XMCD at Co/TNAP interface, while the magnetic moment of Co is slightly reduced by 0.8% from XMCD sum rules. The results indicate that the TNAP adsorption does not degrade the magnetic properties of Co, an advantage in device applications. Thus, interface enginnering by TNAP-induced HIS may provide a feasible way to tune spin selection and manipulate spin injection at FM/OSC interface, which holds the promise to engineer spin injection efficiency in real devices.

**4. Experimental Section**

*Sample fabrication*: The TNAP was purchased from Tokyo Chemical Indusctry (TCI) Europe, while $Alq_3$ and $C_{60}$ were both commercial products from Sigma-Aldrich. All chemicals were ready for use whithout purification, but they were all degassed in UHV chamber before real deposition. Au-coated Si wafer was used as the substrate, and it was cleaned by Ar sputtering for 20 minutes before the metal deposition and confirmed by UPS and XPS measurements. Co was deposited on clean Au surface by use of an UHV e-beam evaporator (Omicron EFM3) at a deposition rate of about 3 Å / min. The thickness of Co film was about 5~6 nm. TNAP was *in-situ* evaporated on clean Co film from a simple Knudsen cell with a flux rate of about 2 Å/min. Thick TNAP film was heated for 5 minutes at 150° in UHV chamber, finally ~ML (0.8 nm) TNAP can be obtained on Co substrate, which was confirmed by the constant C/Co ratio in XPS. For trilayer heterojunctions, Alq$_3$ and C$_{60}$ were deposited on ~ML TNAP from Knudsen cells at a rate of 2 Å/min. All thickness was estimated from the attenuation of the core level signals of the bottom layer (Au 4f), and all samples are *in-situ* fabricated and immediately transferred to analysis chamber for photoelectron spectroscopy.

*Photonelectron spectroscopy*: The XPS and UPS experiments were carried out using a Scienta ESCA 200 spectrometer. The vacuum system consists of an analysis chamber and preparation chamber. XPS and UPS measurements were performed in the analysis chamber at a base pressure of $10^{-10}$ mbar, using monochromatized Al (K$_\alpha$) X-rays at hν=1486.6 eV and He I radiation at hν=21.2 eV, respectively. The experimental conditions were such that the full width at half maximum (FWHM) of the Au 4f7/2 line was 0.65 eV. The binding energies were obtained referenced to the Fermi level with an error of ±0.1 eV. Sputtering and material depositions were done in a preparation chamber with a base pressure of $10^{-9}$ mbar. The take-off angle noted in the figure is defined as the angle between the direction of the detected electrons and the surface of the sample, i.e., a 90° take-off angle means that the electrons are detected leaving perpendicular to the surface (parallel to the surface normal).

*Synchrotron experiments*: Both XAS and XMCD were measured at room temperature. XMCD spectra were obtained in remanence, by taking the difference between XAS spectra recorded with opposite in-plane magnetization directions. The samples were magnetized by applying an in-plane magnetic field pulse of 300 Oe. The angle of incidence of the photon beam was set to 10° relative to the sample normal, and the degree of circular polarization is 85%. All the XAS and XMCD measurements were normalized to the incident photon flux using the TEY of a gold grid, on which a fresh layer of gold was deposited prior to the measurements.

*Theoretical calculation:* The geometry of TNAP molecular was optimized at the unrestricted B3LYP/6-31G* level. The gradient-corrected Becke[46] (BE88) exchange functional and the Perdew[47] (PD86) correlation functional were used to calculate the NEXAFS spectra with the full core hole (FCH) approach using ΔKohn-Sham (ΔKS) scheme, wherein a particular x-ray transition energy is obtained as the difference between the energy of the exited state and that of the ground state. The spectra calculations were carried out at the density functional theory level with the StoBe package[48]. The orbital basis set used were IGLO-III for the excited atom and triple-ζ valence for the rest. Miscellaneous auxiliary basis sets were also set for all atoms, and an effective core potential (ECP) was used to help the convergence of core-hole state. A Gaussian function with a full-width-at-half-maximum (FWHM) of 0.2 eV was used to convolute the spectra below the ionization potential (IP), while a Stieltjes imaging approach was used to describe the spectra above the IP in the continuum[49]. Finally, a relativistic effects correction term of 0.3 eV for N and 0.2 eV for C which associated with the removal of one electron is applied.

**Acknowledgements**
This work is funded through the European Union Seventh Framework Programme (FP7/2007-2013) under grant agreement no. 263104. In general, the Surface Physics and Chemistry division is supported by the Swedish Research Council (project grant and Linneus center) and the Knut and Alice Wallenberg Foundation. X. Li and Y. Luo acknowledge Goran Gustafsson Foundation for Research in Natural Sciences and Medicine, and the Swedish Research Council (VR). The Swedish National Infrastructure for Computing (SNIC) is acknowledged for the computational resources.

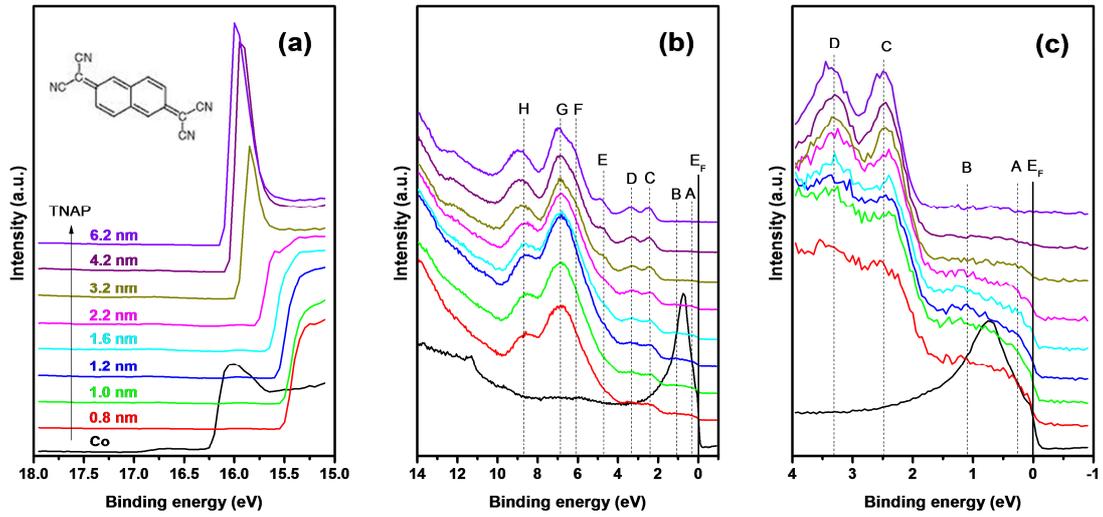

**Figure 1.** UPS spectra of TNAP deposited on Co substrate. (a) Secondary electron cutoff, (b) valence band, and (c) detailed spectral features near $E_F$ with increasing thickness of TNAP. The thickness of TNAP increases from bottom to top as the indication of the arrow in (a), and the insert in (a) shows the chemical structure of TNAP.

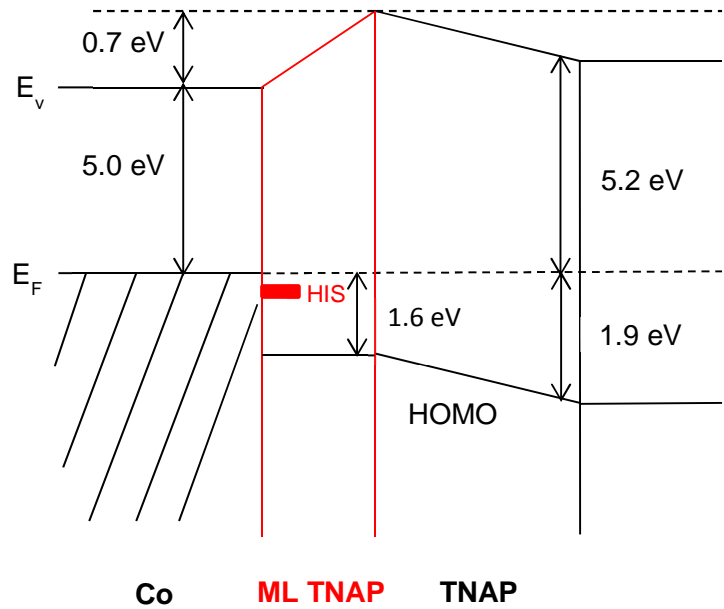

**Figure 2.** The scheme of energy level alignment for TNAP on Co substrate with the thickness from ~ML to bulk, where the red rectangle indicates hybrid interface state (HIS).

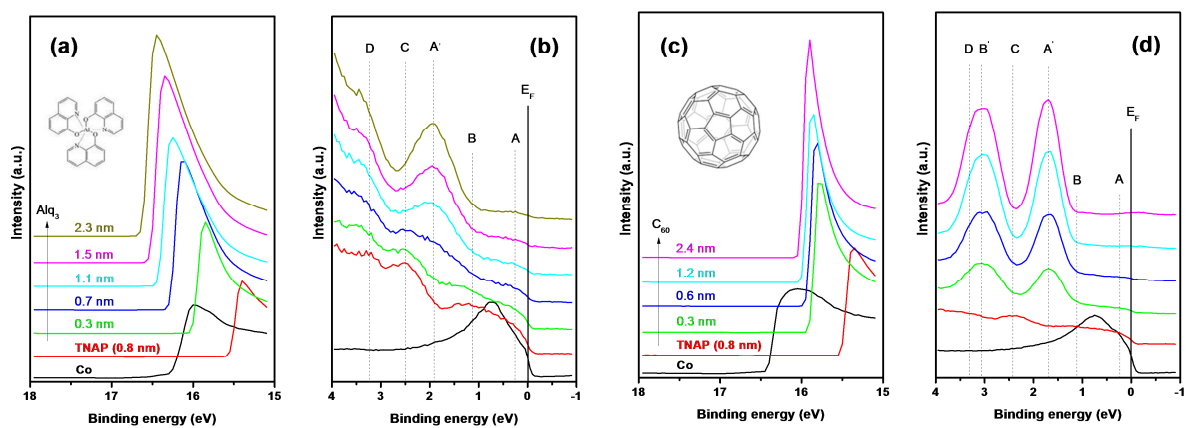

**Figure 3.** UPS spectra of typical heterojunction interfaces. (a) Secondary electron cutoff and (b) valence band near $E_F$ for Co/TNAP/Alq$_3$; (c) Secondary electron cutoff, (d) valence band near $E_F$ for Co/TNAP/C$_{60}$. The arrow in (a) and (c) shows the thickness increasing from bottom to top, and the inserts in (a) and (c) show the chemical structure of Alq$_3$ and C$_{60}$.

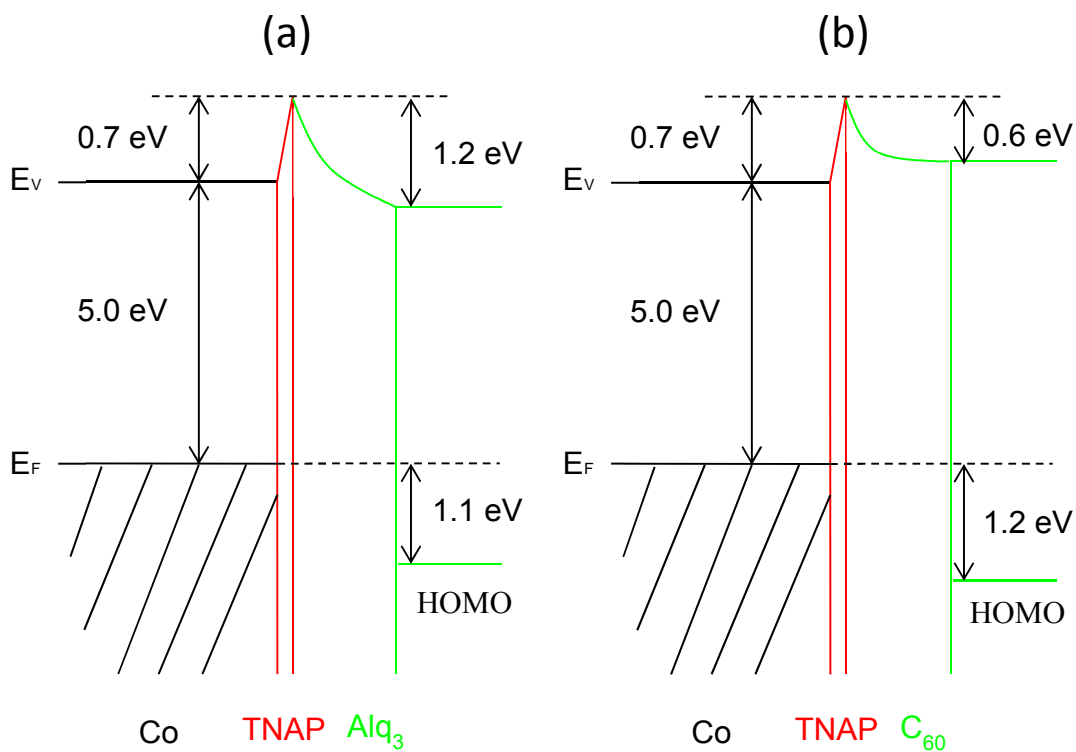

**Figure 4.** The schemes of energy level alignment for different hybrid interfaces. (a) Co/TNAP/Alq$_3$ and (b) Co/TNAP/C$_{60}$.

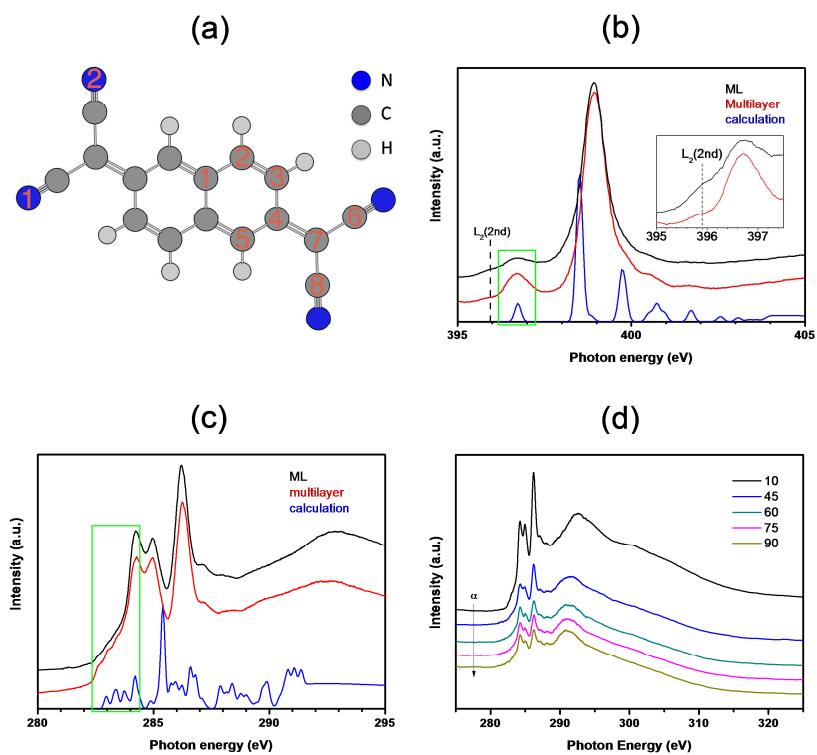

**Figure 5.** NEXAFS spectra for Co/TNAP interface. (a) Optimized structure of TNAP from the calculation, in which the atoms C, N and H are shown in blue, grey and light grey, and the number labeled shows the different atom types; (b) N K-edge and (c) C K-edge of multilayer and ~ML TNAP on Co, and the calculated results for gas-state TNAP are also given as the reference. The green rectangles in (b) and (c) show the evolution of frontier features, and the inserted in (b) shows the Co $L_2$ (2nd) peak position from the partial zoom of the curves; (d) angle-dependence NEXAFS for C K-edge in ~ML TNAP on Co, in which $α$ means the angle between the incident X-ray and sample surface, and the arrow indicates $α$ increasing from 10 to 90 degree.

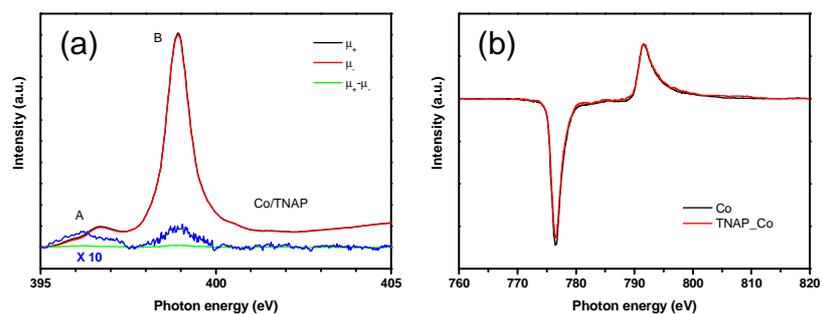

**Figure 6.** XMCD spectra in TEY mode for Co/TNAP interface. (a) NEXAFS and XMCD results in N K-edge for Co/TNAP interface, in which A and B indicate two peaks for XMCD signals; (b) Co L-edge XMCD results before and after adsorption of TNAP on Co.

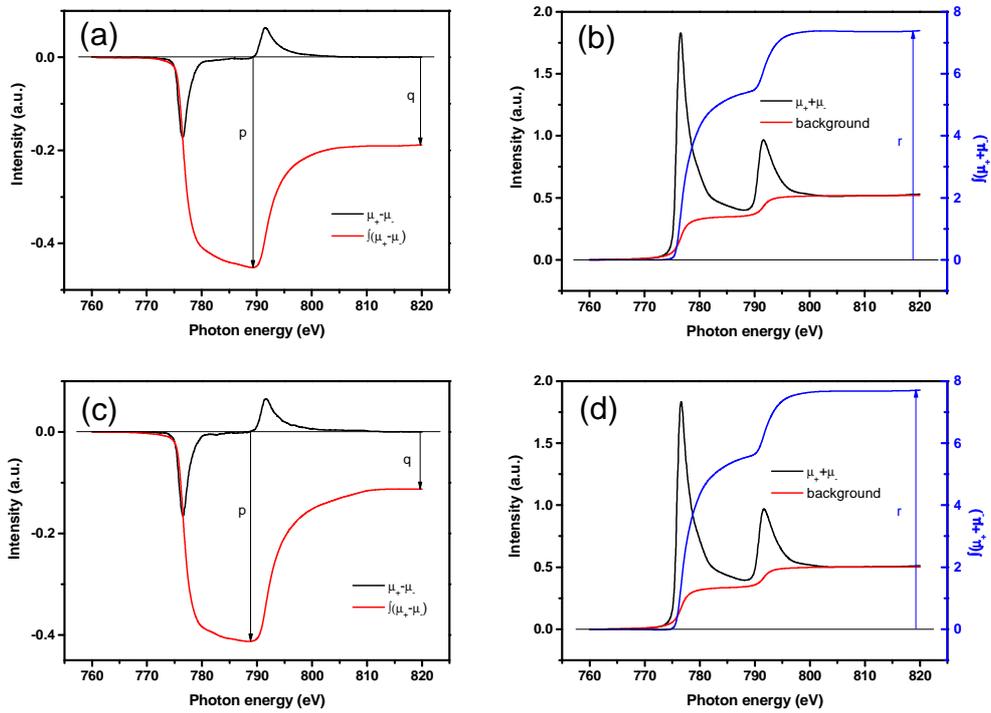

**Figure 7.** Sum rule analysis of Co L-edge NEXAFS and XMCD spectra. (a) XMCD and (b) summed NEXAFS spectra and their integrations for Co substrate, (c) XMCD and (d) summed NEXAFS spectra and their integrations for TNAP adsorption on Co.

**Table 1.** Orbital and spin magnetic moments of Co before and after TNAP adsorption in units of $\mu_B$/atom from sum rules.

| parameters | $m_{spin}$ | $m_{orb}$ | $m_{orb}/m_{spin}$ |
|---|---|---|---|
| Co | 0.788 | 0.101 | 0.128 |
| Co/TNAP | 0.782 | 0.058 | 0.074 |

# Supplementary

**Hybrid interface states (HIS)**

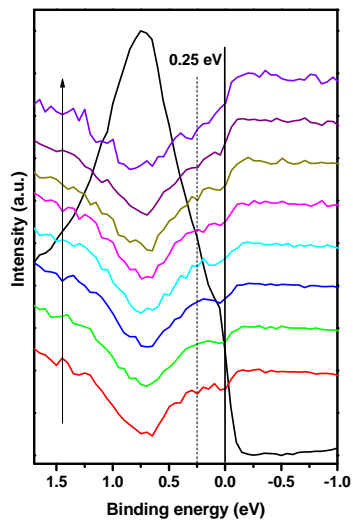

Figure S1. The normalized spectra feature at 1.7 eV below $E_F$ with the TNAP thickness from ~ML to multilayer range, in which the contribution from the Co substrate was subtracted. The TNAP thickness increases from bottom to top as the indication of the arrow.

For the UPS spectra of Co/TNAP interfaces, if we check the detailed information in valence band region near the $E_F$ (Figure S1), and we subtract the contribution from the Co substrate in all spectra, there is a new peak around 0.25 eV formed below $E_F$, and its intensity gradually decreases when the thickness of TNAP increases from ~ML to thick, and finally it disappears for bulk TNAP, hence this peak is assigned as HIS caused by the interaction between Co and TNAP, and it shows good agreement with our ananlysis on the spectral feature (peak A) in the valence band. Here we have to point out that the HIS peak exists not only in ~ML TNAP, but also over a thickness range. As we have discussed in the main text, ~ML can't cover all Co surface, and then there is a thickness range for TNAP molecules to fully cover the bare Co surface. Although we use He I (21.2 eV) radiation source here, it shows a typical HIS as the reports by S. Lach

before[S1]. From Figure S1, we also find the HIS peak shows a broad one from 0.05 to 0.5 eV below $E_F$, which can be an evidence of strong interaction between Co and TNAP. Here all spectra were normalized before the subtraction with the contribution from Co, and the energy range (1.7 eV below $E_F$) was chosen because in this region DOS from pristine TNAP is not expected.

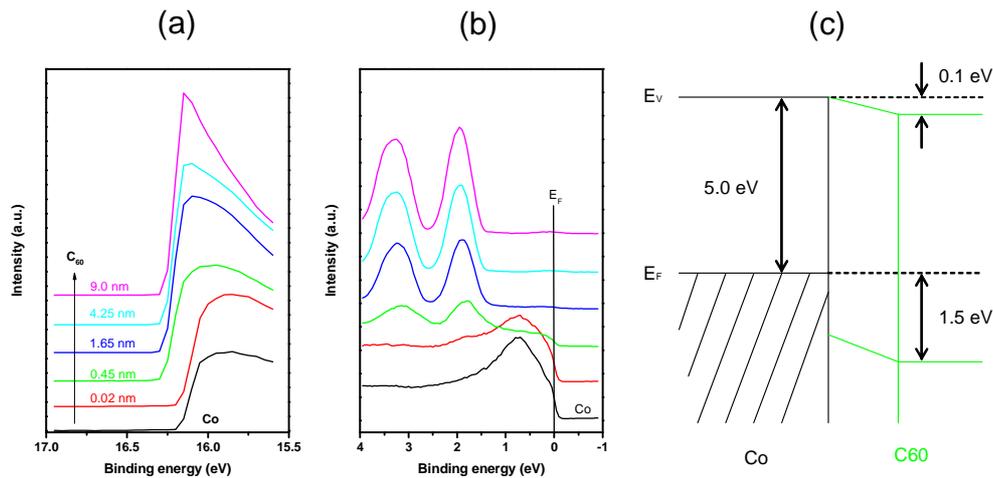

Figure S2. UPS spectra of $C_{60}$ deposited on Co substrate. (a) Secondary electron cutoff, (b) detailed spectraL features near $E_F$ with increasing thickness of TNAP, and (c) Energy level alignment of Co/$C_{60}$ interface. The TNAP thickness increases from bottom to top as the indication of the arrow in (a).

### Energy level alignment at Co/$C_{60}$ interface

In Figure S2, we show the UPS spectra of $C_{60}$ deposited on Co substrtae. As the thickness of $C_{60}$ increases from ~ML to bulk, there is very little reduction in work function, with totally 0.1 eV downshift compared with that of Co. The HOMO edge is situated at 1.5 eV and the valence band is very stable in the binding energy position, only the intensity increases gradually. The energy level alignment is shown in Figure S2c as the reference for Co/TNAP/$C_{60}$ trilayer interface in the main text.

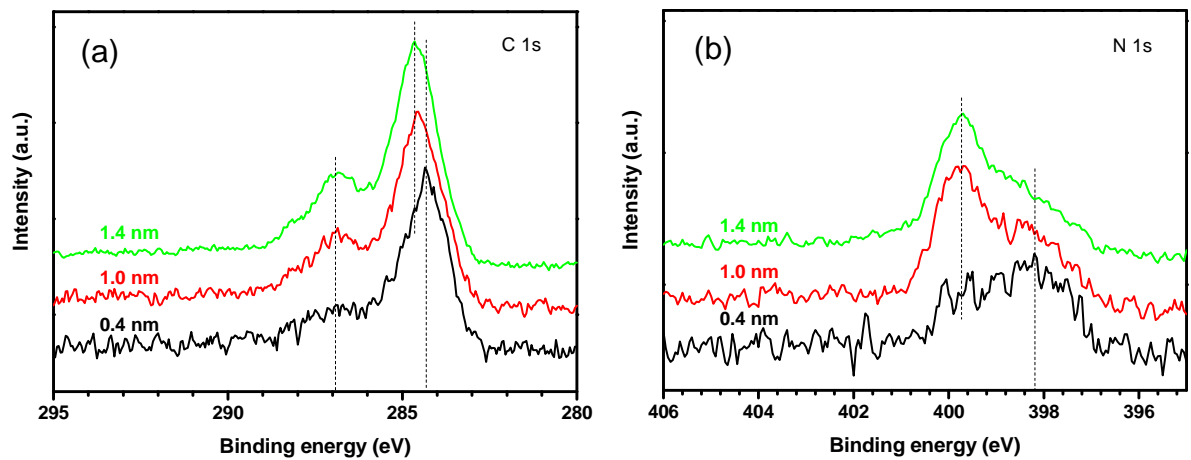

Figure S3. XPS spectra for C 1s and N 1s core levels of TNAP as a function of its thickness on Co.

**XPS analysis at the Co/TNAP interface**

We have also obtained XPS spectra for C 1s and N 1s core levels of TNAP as a function of its thickness on Co (Figure S3), by which to determine the bonding between Co and TNAP molecules.

For C 1s in Figure S3a, the main feature is situated at 284.3 eV when a ~ML TNAP (0.4 nm) is deposited on Co, and it slightly shifts to higher binding energy and becomes stable at 284.65 eV with subsequent deposition of TNAP molecules, which then may be attributed to the charge transfer from Co to TNAP at the interface. There is still another peak around 286.9 eV, attributed to the C in cyano group, and its intensity is gradually enhanced as the thickness increases. For N 1s in Figure S3b, the similar peak evolution can be found. A distinct peak around 398.15 eV exists in ~ML TNAP on Co, and its intensity is reduced when more TNAP is adsorbed on the surface and disappears at a large thickness of TNAP. Subsequently, the low binding energy peak is assigned to bonding between Co and TNAP at the interface through the nitrogen atoms, which reults in the HIS at the interface. The main peak for N 1s appears around 399.75 eV from –CN group as the intensity gradually becomes stronger when the thickness increases. Based on the discussion on the XPS spectra of C 1s and N 1s, the chemical interaction between Co and TNAP molecule probably happens on N atoms in -CN group, which confirms the hybridization in NEXAFS and supports the finding of induced magnetization in XMCD for N K-edge.